\documentclass[aps,prl,twocolumn,showpacs,groupaddress]{revtex4}

\usepackage{amsmath}
\usepackage{amssymb}
\usepackage{epsfig}
\usepackage{bm}
\usepackage{longtable}
\usepackage{graphicx}
\usepackage{graphics}

\begin{document}

\title{
Friedel sum rule for an interacting multi-orbital quantum dot
}

\author{Massimo Rontani}
\email{rontani@unimore.it}
\homepage{www.nanoscience.unimo.it/max_index.html}
\affiliation{CNR-INFM National Research Center S3, 
Via Campi 213/A, 41100 Modena, Italy}

\date{\today}

\begin{abstract}
A generalized Friedel sum rule is derived for a 
quantum dot with internal orbital and spin degrees of freedom. 
The result is valid when all many-body
correlations are taken into account and it links the phase shift 
of the scattered electron to the displacement
of its {\em spectral} density into the dot.
\end{abstract}


\maketitle
The Friedel sum rule (FSR) is one of the few exact results of 
solid state physics \cite{Grosso}, with a vaste range of applications in
the fields of scattering theory, magnetic and non magnetic impurities
in metals \cite{Grosso}, Kondo effect \cite{Hewson}, 
and, very recently, coherent transport
through quantum dots (QDs) \cite{Hackenbroich01} and molecules \cite{Datta97}. 
This powerful relation connects the charge density $\varrho(\omega)$
displaced by an impurity or a nano-object, acting as a 
scattering center in a conductor,
with the variation of the phase shift $\delta_{\text{F}}$ of the 
scattered wave
with respect to the energy $\hbar\omega$:
\begin{equation}
\frac{1}{\hbar}\frac{d \delta_{\text{F}}(\omega)}{d\omega} = \frac{\pi}{e}  
\varrho(\omega).
\label{eq:start}
\end{equation}
The Friedel phase $\delta_{\text{F}}$ of Eq.~(\ref{eq:start}),
which holds for single-channel elastic scattering only, is
linked to the eigenvalue of the scattering matrix $S$
through the identity
$S=\text{e}^{2\text{i}\delta_{\text{F}}}$. In the many-channel case, 
$\delta_{\text{F}}$ appearing on the left hand 
side of (\ref{eq:start}) is replaced by $\text{Tr}\,\text{ln}\,S/
2\pi \text{i}$ \cite{Langer61}. 

Recently Lee \cite{Lee99} and Taniguchi and B\"uttiker \cite{Taniguchi99} 
showed the relevance of FSR in
measurements of the transmission phase acquired 
by an electron passing through a QD embedded in the 
arm of an Aharonov-Bohm interferometer \cite{Schuster97}.
These experiments \cite{Schuster97,Moty} allow for
both directly measuring the phase shift and 
arbitrary controlling the Fermi
energy (or, equivalently, the plunger gate voltage of
the QD), namely the two quantities appearing on both sides of
(\ref{eq:start}). 
Even if the identification of $\delta_{\text{F}}$ with
the transmission phase is unjustified in generic situations
\cite{Lee99,Taniguchi99},
nevertheless Aharonov-Bohm interferometry paves the way to the 
direct experimental test of
FSR for a QD whose internal structure, charge, spin, and correlation
can be externally controlled.  
Indeed, the ``non-universal'' behavior of the QD transmission phase
in the regime of very few electrons ($N<10$) suggests
that the QD orbital and spin degrees of freedom may play a
major role \cite{Moty}. 

In real QDs used in interference experiments single-particle levels have
a small energy separation ($\sim 0.5$ meV), if compared to characteristic
Coulomb energies ($\sim 1-3$ meV) \cite{Moty},
and therefore many of them should be included in any reliable
model for electron correlation \cite{Garcia05,Rontani06}.
The FSR is generally  believed to hold even 
in the presence of electron-electron interaction \cite{Lee99}. 
This was rigorously demonstrated only in two cases: 
(i) electrons in the metal 
form a Fermi liquid and the impurity has no internal degrees
of freedom \cite{Langer61} (ii) the interaction is limited to a localized 
impurity orbital (Anderson model \cite{Langreth66} and its specific extensions
to open atomic shells \cite{Langreth66,Hewson}). 
This Letter shows that the FSR must be reformulated in the experimentally 
relevant case of a multi-level interacting dot. 
A generalized statement holds, with the spectral density
$\mathcal{N}(\omega)$ of the scattered electron accumulated in the
QD replacing the non-interacting density of states $\varrho(\omega)/e$
appearing in (\ref{eq:start}). We discuss the 
relevance of this result for the qualitative understanding of a few
puzzling features of the experiment of Ref.~\onlinecite{Moty}.   

We consider a generic system whose Hamiltonian $H$ is made 
of three terms separately describing
the multi-orbital interacting quantum dot, $H_{\text{dot}}$,
the conduction electrons in the leads, $H_{\text{lead}}$, and the
hopping term between dot and leads, $H_{\text{mix}}$:  
\begin{equation}
H = H_{\text{dot}} + H_{\text{lead}} + H_{\text{mix}}.
\end{equation}
The conduction electrons, in typical experimental setups,
can freely move in a two-dimensional heterostructure
\cite{twoleads},
according to the Hamiltonian
$H_{\text{lead}} = \sum_{{\bm k}\sigma}
\varepsilon_{\bm k}\, n_{{\bm k}\sigma}$,
where $n_{{\bm k}\sigma}=c^{\dagger}_{{\bm k}\sigma}c_{{\bm k}\sigma}$
and $c^{\dagger}_{{\bm k}\sigma}$ creates an electron into
the Bloch state of crystal momentum
${\bm k}$,  spin $\sigma$, and energy $\varepsilon_{\bm k}$. 
The QD Hamiltonian $H_{\text{dot}}$ includes both the single-particle
and the interacting part:
\begin{equation}
H_{\text{dot}} = \sum_{\alpha i \sigma}
\varepsilon_{\alpha i} n_{\alpha i \sigma} + H_{\text{int}}.
\label{eq:Hqd}
\end{equation}
In order to label the QD orbitals, we here introduce two indices, 
$\alpha=1\ldots, N_{\text{class}}$ and $i=1,\ldots, N_{\alpha}$, 
respectively. The former index, $\alpha$,
labels the $N_{\text{class}}$ irreducible
representations of the QD point-symmetry group, while
the latter, $i$, enumerates the truncated set of $N_{\alpha}$ orbitals 
considered, belonging to the same $\alpha$ 
representation. 
We assume that QD orbitals form also 
a basis for representing the symmetry of the
whole dot + leads system \cite{Klein66}.
Typical symmetry groups of realistic devices range from 
$D_{\infty h}$ to $C_{2v}$, going from circular \cite{Garcia05} to 
elliptic \cite{Zunger05} dots, respectively \cite{notasimmetria}. 
In Eq.~(\ref{eq:Hqd}) $n_{\alpha i \sigma}=
c^{\dagger}_{\alpha i \sigma}c_{\alpha i \sigma}$, 
$c^{\dagger}_{\alpha i \sigma}$ creates an electron
with spin $\sigma$ in the orbital $(\alpha,i)$ of energy
$\varepsilon_{\alpha i}$, and $H_{\text{int}}$ includes the
full intra-dot Coulomb interaction. Note that
$H_{\text{int}}$ does not commute with $n_{\alpha i \sigma}$ except if
there is either one level only, $N_{\text{class}}=1$, $N_{\alpha}=1$ 
(non degenerate Anderson model), or the Coulomb interaction
takes an oversimplified form.
Here the full inclusion of all Coulomb
matrix elements in $H_{\text{int}}$
turns out to be crucial in the following.
Finally, the tunneling term $H_{\text{mix}}$ allows for electron 
hopping between delocalized Bloch states $\bm{k}$ and 
confined orbitals $(\alpha,i)$
via the matrix elements $V_{\bm{k}\alpha i}$:
\begin{equation}
H_{\text{mix}} = \sum_{{\bm k}\alpha i \sigma}\left(
V_{\bm{k}\alpha i} c^{\dagger}_{{\bm k}\sigma}
c_{\alpha i \sigma} \quad+\text{c.c.}\right).
\end{equation} 

As a first step,
we introduce the zero-temperature exact retarded Green's function 
\begin{equation}
\text{i}\mathcal{G}_{XX^{\prime}}(t)
=\vartheta (t)\left<c_X(t)c^{\dagger}_{X^{\prime}}
(0) + c^{\dagger}_{X^{\prime}}(0)c_X(t)\right>,
\label{eq:G}
\end{equation}
where either $X=\alpha i \sigma$ or $X=\bm{k}\sigma$, 
and $\left<\ldots\right>$ is the average 
on the interacting ground state $\left|\Psi_0\right>$ of the
whole dot + leads system in the Heisenberg representation. 
We work with the analytic continuation of
the Fourier transform of (\ref{eq:G}) in the complex plane of
the energy, $\mathcal{G}_{XX^{\prime}}(z)$. 

In the following we neglect
spin-flip scattering processes \cite{Langreth66}, 
and therefore we may drop spin indices of $\mathcal{G}$. 
Moreover, off-diagonal Green's functions of 
type $\mathcal{G}_{\alpha i \beta j}(z)$ must vanish 
due to symmetry [but not
$\mathcal{G}_{\alpha i \alpha j }(z)$, henceforth indicated as
$\mathcal{G}^{\alpha}_{ i j }(z)$].
We first consider the non-interacting case (then  
$H_{\text{int}}=0$ and $G$ is printed in italic), 
where explicit solutions are readily 
obtained using the Green's function's equations of motions 
\cite{Anderson61,Klein66}:
\begin{equation}
\hbar^{-1}G_{\bm{k k}}(z) = \frac{1}{ z - \varepsilon_{\bm k} }
+ \sum_{\alpha ij}\frac{ V_{\bm{k} \alpha i} V_{\alpha j \bm{k} } }
{\left(z -\varepsilon_{\bm k}\right)^2}G^{\alpha}_{ i j}(z),
\label{eq:Gkk}
\end{equation}
and
\begin{equation}
\sum_{m}\left[ \left(z-\varepsilon_{\alpha i}\right)\delta_{im}-
\Delta^{\alpha}_{ i m}(z)\right]G^{\alpha}_{ m j}(z)=\hbar\,\delta_{ij},
\label{eq:matrix1}
\end{equation}
with $\Delta^{\alpha}_{ i j}(z)$ being the self-energy due to
the dot-lead interaction:
$\Delta^{\alpha}_{ i j}(z) = \sum_{\bm k} V_{\alpha i \bm{k}}
V_{\bm{k} \alpha j} ( z - \varepsilon_{\bm k} )^{-1}$.
In the non-interacting case the 
diagonal QD Green's function assumes
the familiar form
$\hbar^{-1}G^{\alpha}_{ i i}(z) = \left[ z - \varepsilon_{\alpha i} - 
\Delta^{\alpha}_{ i i}(z)\right]^{-1}$,
where $\text{Im}\left[\Delta^{\alpha}_{ i i}(z)\right]$ is the virtual
level width, and $\text{Re}\left[\Delta^{\alpha}_{ i i}(z)\right]$ 
renormalizes the single-particle level $\varepsilon_{\alpha i}$.
Equation (\ref{eq:matrix1}) is modified to take into account
electron-electron interaction by introducing the intra-dot
proper self-energy matrix $\Sigma^{\alpha}_{ i j}(z)$:
\begin{equation}
\sum_{m}\left[ \left(z-\varepsilon_{\alpha i}\right)\delta_{im}
-\Sigma^{\alpha}_{ i m}(z)-
\Delta^{\alpha}_{ i m}(z)\right]\mathcal{G}^{\alpha}_{ m j}(z)=
\hbar\,\delta_{ij}.
\label{eq:matrix2}
\end{equation}

Henceforth we focus on the fully interacting system, and 
we consider the case of elastic scattering 
when only a single QD level, $(\bar{\alpha},\bar{\i})$, 
is coupled to the leads. Indeed, this is a reasonable scenario
for the electrostatic potential barriers
separating dot and leads in many experimental setups \cite{Moty}, 
where matrix elements $V_{\bm{k} \alpha i}$
strongly depend on both energies 
$\varepsilon_{\bm k}$ and $\varepsilon_{ \alpha i}$, respectively.
We then set $V_{\bm{k} \alpha i}=0$ if $(\alpha, i) \neq (\bar{\alpha},
\bar{\i})$ and  
$V_{\bm{k} \bar{\alpha}\bar{\i}}\neq 0$ 
with $\varepsilon_{\bar{\alpha}\bar{\i}} 
- \varepsilon_{\text{cut}}
\le \varepsilon_{\bm k} \le \varepsilon_{\bar{\alpha}\bar{\i}}
+ \varepsilon_{\text{cut}}$, 
where $\varepsilon_{\text{cut}}$ is a suitable cutoff.
Note that Coulomb correlation is included in full
and the intra-dot self-energy has off-diagonal matrix elements,
$\Sigma^{\alpha}_{ij}(z)\neq 0$, $\alpha = 1,\ldots,N_{\text{class}}$,
$i,j=1,\ldots,N_{\alpha}$.
This scenario is generic enough to correctly describe
many experimental situations, except the case of degeneracies 
between QD levels $\varepsilon_{\bar{\alpha}\bar{\i}}$ and 
$\varepsilon_{\beta j}$.

We proceed in close analogy with Langreth \cite{Langreth66} and  
calculate the charge $\mathcal{N}_{\bar{\alpha}\bar{\i}}$ 
(in units of $e$) displaced at the dot 
level $(\bar{\alpha},\bar{\i})$ 
as the difference between the charge at equilibrum in the presence and 
in the absence of the dot, respectively:
\begin{eqnarray}
\mathcal{N}_{\bar{\alpha}\bar{\i}} &=& -\frac{1}{\hbar\pi}\text{Im}
\int_{-\infty}^{\mu+\text{i}\eta}dz
\Big\{\Big[ \sum_{\bm k}
\mathcal{G}_{\bm{k k}}(z)
+\mathcal{G}^{\bar{\alpha}}_{\bar{\i}\bar{\i}}(z) \Big] \nonumber\\
&&-\sum_{\bm k} G^{\,\text{free}}_{\bm{k k}}(z)\Big\} ,
\label{eq:N}
\end{eqnarray}
where $\eta$ is a positive infinitesimal quantity,
$\mu$ is the equilibrium chemical potential \cite{Abrikosov},
and $\hbar^{-1}
G^{\,\text{free}}_{\bm{k k}}(z)=(z - \varepsilon_{\bm k})^{-1}$
is the propagator of a free traveling wave in the absence of the QD.
By using Eqs.~(\ref{eq:N}) 
and (\ref{eq:Gkk}) where $G$'s are replaced with $\mathcal{G}$'s, 
one obtains the following expression for $\mathcal{N}_{\bar{\alpha}\bar{\i}}$:
\begin{equation}
\mathcal{N}_{\bar{\alpha}\bar{\i}} = -\frac{1}{\hbar\pi}\text{Im}
\int_{-\infty}^{\mu+\text{i}\eta}dz 
\left[ 1 - \frac{\partial 
\Delta^{\bar{\alpha}}_{\bar{\i}\bar{\i}}(z) }{\partial z} \right]
\mathcal{G}^{\bar{\alpha}}_{\bar{\i}\bar{\i}}(z).
\label{eq:partial}
\end{equation}
We now use the identity
\begin{equation}
\text{Im}\int_{-\infty}^{\mu+\text{i}\eta}dz
\frac{\partial\Sigma^{\bar{\alpha}}_{\bar{\i}\bar{\i}}(z)}{\partial z}
\mathcal{G}^{\bar{\alpha}}_{\bar{\i}\bar{\i}}(z) = 0,
\label{eq:Luttinger}
\end{equation}
which is a natural generalization of the
Luttinger relation of Fermi liquids \cite{Luttinger60}, 
and it has been already applied by 
Langreth \cite{Langreth66} to the single-level
case. 
%[cf.~also Refs.~\cite{Potthoff} 
%and \cite{Sachdev} for extensions to the multi-level case
%similar to (\ref{eq:Luttinger})].
Combining Eqs.~(\ref{eq:Luttinger}) and (\ref{eq:partial}) gives
\begin{equation}
\mathcal{N}_{\bar{\alpha}\bar{\i}} 
= \frac{1}{\pi}\text{Im}\int_{-\infty}^{\mu+\text{i}\eta}dz
\frac{\partial}{\partial z} \text{ln}\,
\mathcal{G}^{\bar{\alpha}}_{\bar{\i}\bar{\i}}(z).
\label{eq:log}
\end{equation}
Since the asymptotic form of the Green's function as $z\rightarrow 
-\infty$ is $\mathcal{G}^{\alpha}_{ij}(z)\approx \hbar\delta_{ij}/z$,
Eq.~(\ref{eq:log}) may be casted into the form
\begin{equation}
\mathcal{N}_{\bar{\alpha}\bar{\i}} = \frac{1}{\pi}\text{Im}\left[\text{ln}\,
\mathcal{G}^{\bar{\alpha}}_{ \bar{\i}\bar{\i} }
(\mu +\text{i}\eta) - \text{i}\pi\right],
\label{eq:almost}
\end{equation}
where we have chosen the cut along the positive real axis. 
To connect the displaced charge $\mathcal{N}_{\bar{\alpha}\bar{\i}}$ 
to the Friedel phase, we observe that the T scattering matrix of
the single channel $\bm{k}$ is 
%which is the argument of the T scattering matrix \cite{Hewson}, 
%observe that Eq.~(\ref{eq:Gkk}) can be written as
%\begin{equation}
%\mathcal{G}_{\bm{k k}}(z) = G^{\,\text{free}}_{\bm{k k}}(z)
%+ G^{\,\text{free}}_{\bm{k k}}(z) \left|V_{\bm{k}\bar{\alpha} \bar{\i}} 
%\right|^2 
%\mathcal{G}^{\bar{\alpha}}_{ \bar{\i}\bar{\i}}(z)
%G^{\,\text{free}}_{\bm{k k}}(z).
%\label{eq:Gkkbis}
%\end{equation}
%From Eq.~(\ref{eq:Gkkbis}) one identifies the T scattering matrix of 
%the single channel $\bm{k}$ as 
$\mathcal{T}_{\bm{k k}}(z) =
\left|V_{\bm{k} \bar{\alpha} \bar{\i}} \right|^2\hbar^{-1}
\mathcal{G}^{\bar{\alpha}}_{ \bar{\i}\bar{\i}}(z)$ (cf.~Sec.~5.2 of 
Ref.~\onlinecite{Hewson}), therefore the phase on
the energy shell $\delta_{\text{F}}(\mu/\hbar)$ is
\begin{equation}
\delta_{\text{F}}(\mu/\hbar) = 
\text{arg}\,\mathcal{T}_{\bm{k k}}(\mu+\text{i}\eta)
=\text{Im}\,\text{ln}
\mathcal{G}^{\bar{\alpha}}_{ \bar{\i}\bar{\i}}(\mu+\text{i}\eta)-\pi,
\label{eq:Friedeldef}
\end{equation}
where we pick the branch of the logarithm as before
and add the reference constant $-\pi$,
so that $0\le \delta_{\text{F}} \le \pi$.
Comparison of (\ref{eq:almost}) and (\ref{eq:Friedeldef}) gives
the desired result:
\begin{equation}
\delta_{\text{F}}(\mu/\hbar) = \pi\, \mathcal{N}_{\bar{\alpha}\bar{\i}}.
\label{eq:Friedel}
\end{equation}

The exact sum rule (\ref{eq:Friedel}) states that the Friedel phase
shift, due to scattering by the dot, is proportional 
to the net charge accumulated at the dot level $(\bar{\alpha},\bar{\i})$
with respect to the free system (i.e.~without dot). In the case of a single
level ($N_{\text{class}}=1$, $N_{\alpha}=1$), Eq.~(\ref{eq:Friedel})
is equivalent to the result of Langreth for the 
Anderson model \cite{Langreth66}. 
In the many-level case, (\ref{eq:Friedel}) is a non trivial 
generalization of previous theories 
(see e.g.~\cite{Langreth66} and \cite{Hewson}), which can be 
summarized as follows:
The sum of phase shifts due to consecutive filling of many
levels is fixed by the {\em total} charge which can be placed into the 
dot, ruled by $\mu$. If the dot levels, possibly broadened
due to hybridization with continuum states, lie well below $\mu$,
and $T>T_{\text{K}}$, where $T_{\text{K}}$ is the Kondo
temperature,
then the afore mentioned charge is an integer quantity fixed by orbital
degeneracy.
According to (\ref{eq:Friedel}), however, this simple picture is 
generally incorrect, as the following conceptual tunneling
experiment illustrates.

Think of varying continuously the chemical potential $\mu$
across an energy window centered around the resonant value
$\hbar\omega_{\text{res}}$, 
which is implicitely given by the real part of a certain pole of 
$\mathcal{G}^{\bar{\alpha}}_{\bar{\i} \bar{\i}}(z)$,
$\hbar\omega_{\text{res}} = \varepsilon_{\bar{\alpha}\bar{\i}}
+ \text{Re} \left[ \Sigma^{\bar{\alpha}}_{\bar{\i} \bar{\i}}
(\hbar\omega_{\text{res}}) + \Delta^{\bar{\alpha}}_{\bar{\i} \bar{\i}}
(\hbar\omega_{\text{res}}) \right]$.
For the sake of clarity we here focus on the Coulomb blockade regime only,
despite the fact that sum rule (\ref{eq:Friedel}) applies
to the Kondo regime as well. The level at $\hbar\omega_{\text{res}}$
is located between two blockaded regions with $N$ and $N+1$
electrons in the dot, respectively.
The transport window is chosen to be large enough to fully contain the width
of the QD level, given by $\varepsilon_{\text{cut}}$, but 
narrower than the spacing between neighboring resonant levels.
Therefore, as $\mu$ is swept upward across 
the QD level, the dot is charged by exactly one electron, in
addition to those already localized. Nevertheless, the variation of
the {\em spectral} density $\Delta\mathcal{N}_{\bar{\alpha}\bar{\i}}$ of 
Eq.~(\ref{eq:Friedel}) across the energy window is generally less than one,
and consequently the phase shift increment of the outgoing wave spreading
out from the dot 
at the top of the energy window is less than $\pi$.

To understand it, note that the total number of 
scattering states of the whole dot + leads system, in the energy window 
considered above, must be 
exactly equal to the sum of both free travelling waves in the leads 
and confined states in the dots,
when the two subsystems  are decoupled ($H_{\text{mix}}=0$). 
Therefore, in this case 
$\Delta\mathcal{N}_{\bar{\alpha}\bar{\i}}$
may be calculated by simply integrating the spectral density
of the isolated dot:
\begin{eqnarray}
\Delta \mathcal{N}_{\bar{\alpha}\bar{\i}} &=& -\frac{1}{\hbar\pi}
\text{Im}\int_{\hbar\omega_{\text{res}}-
\varepsilon_{\text{cut}}+\text{i}\eta}^{\hbar\omega_{\text{res}}+
\varepsilon_{\text{cut}}+\text{i}\eta} dz\,
\mathcal{G}^{\bar{\alpha}\,\text{free}}_{\bar{\i}\bar{\i}}(z)\nonumber\\
&&
= \left|\left<\Psi_0^{N+1}|c^{\dagger}_{\bar{\alpha}\bar{\i}}
|\Psi_0^N\right>\right|^2,
\label{eq:exemplum}
\end{eqnarray}
where $\left|\Psi_0^N\right>$ is the exact interacting ground
state of the isolated dot with $N$ electrons. 
The following is clear: (i) $\Delta \mathcal{N}_{\bar{\alpha}\bar{\i}}$
is a positive quantity, 
which can considerably deviate from the unit charge that
could fill in the level $(\bar{\alpha},\bar{\i})$, 
due to the correlation between electrons localized in the dot. 
$\Delta \mathcal{N}_{\bar{\alpha}\bar{\i}}$ can even be zero due to
spin blockade, namely the difference between total spins of 
$\left|\Psi_0^{N+1}\right>$ and $\left|\Psi_0^N \right>$
is not equal to $\pm 1/2$.
(ii) The missing weight
is recovered by integration on the whole spectrum. (iii) 
Equation (\ref{eq:exemplum}) provides the basis for exact numerical 
evalutation of the phase shift, by computing the spectral density 
of a confined system e.g. by means of the full configuration interaction
method \cite{Rontani06,Rontani06b,Rontani05b}.

The discrepancy between total and spectral charge density of
Eq.~(\ref{eq:Friedel}) can be neglected only if intra-dot correlation
effects are absent, namely the ground state of the isolated
dot is a single Slater determinant. 
In such a case, the matrix element appearing in 
Eq.~(\ref{eq:exemplum})
is either one or zero, depending if level $(\bar{\alpha},\bar{\i})$
is filled in or not, respectively, in the $N\rightarrow N+1$
tunneling event.
This occurs in the degenerate Anderson model 
\cite{Hewson,Langreth66,Anderson61,Yeyati99},
where $H_{\text{int}}$ assumes a Hartree-Fock-like form diagonal
in the $(\alpha,i)$ indices.
While such mean-field model is satisfying for magnetic impurities in a bulk
metal or many-electron dots, it breaks down for 
large dots with very few electrons, like those of Ref.~\cite{Moty},
where correlation effects may dominate \cite{Garcia05}, 
driving the ground state of 
the isolated dot even towards the Wigner crystallization regime
\cite{Rontani06}.

The idea behind (\ref{eq:Friedel}) is general,
as we prove in the following case of arbitrary coupling between
leads and QD levels ($V_{\bm{k}\alpha i}\neq 0$ $\forall$ $\bm{k},
\alpha, i$).  
We start from the exact 
relation between Friedel phase and delay time $\tau_{\text{delay}}$ 
of a traveling wave packet when trapped in the dot \cite{Smith60}:
\begin{equation}
2\frac{d\, \delta_{\text{F}}(\omega)}{d \omega} = 
\tau_{\text{delay}}(\omega),
\label{eq:delay}
\end{equation}
where $\hbar\omega$ is the average energy of the wave packet.
We assume that Eq.~(\ref{eq:delay}), proved for independent particles
\cite{Wigner55},
holds even in the presence of correlation since 
$H_{\text{int}}\neq 0$ only in the region 
$\Omega$ occupied by the dot, while in the outer space the packet
moves freely.
The delay $\tau_{\text{delay}}$
can be calculated as the difference between the 
(``dwell'') times spent by the
packet in $\Omega$ in the presence and in the absence of 
the dot, respectively \cite{Smith60,Hauge89}.

By extending the approach of Iannaccone \cite{Iannaccone95},
we write the wave packet $\left|\Phi\right>$ as a superposition
of the interacting incoming scattering states $\left|\Psi_{\bm{k}}\right>$,
$\left|\Phi\right>=\sum_{\bm{k}} 
\alpha(\bm{k})\left|\Psi_{\bm{k}}\right>$,
where $\sum_{\bm{k}}
\left|\alpha(\bm{k})\right|^2 = 1$ so that
$\left|\Phi\right>$ is normalized to unity.
The exact stationary interacting state, $\left|\Psi_{\bm{k}}\right>$,
satisfies $H \left|\Psi_{\bm{k}}\right>=(E_0 +\varepsilon_{\bm{k}})
\left|\Psi_{\bm{k}}\right>$, with
$E_0$ being the energy of the ground state without the
extra electron to be scattered, and $H\left|\Psi_0\right>=
E_0\left|\Psi_0\right>$. 
%may be written as \cite{Langreth66}:
%\begin{equation}
%\left|\Psi_{\bm{k}}\right> = c^{\dagger}_{\bm{k}}\left|\Psi_0\right>
%+\frac{1}{E_0 +\varepsilon_{\bm{k}}+\text{i}\eta-H}H_{\text{mix}}\,
%c^{\dagger}_{\bm{k}}\left|\Psi_0\right>.
%\label{eq:LS1}
%\end{equation}
The probability amplitude for finding the scattered electron
at position $\bm{r}$ is obtained by projecting $\left|\Phi\right>$
on an eigenstate of position of the extra electron
\cite{Rontani05b}, $\Psi^{\dagger}(\bm{r},t)
\left|\Psi_0\right>$, where we introduce the field operator 
$\Psi^{\dagger}(\bm{r},t)$, creating an electron
in $\bm{r}$. The field $\Psi(\bm{r},t)$
may be decomposed onto a mixed basis of
Bloch free waves $\phi_{\bm{k}}(\bm{r})$ and confined QD orbitals
$\phi_{\alpha i}(\bm{r})$: $\Psi(\bm{r},t)=\sum_{\bm{k}}\phi_{\bm{k}}
\!(\bm{r})\,c_{\bm{k}}+\sum_{\alpha i}\phi_{\alpha i}\!(\bm{r})\,
c_{\alpha i}$ \cite{orthogonality}.
We therefore 
define the mean dwell time in $\Omega$ associated with the wave packet 
$\left|\Phi\right>$ as 
\begin{equation}
\int_{-\infty}^{\infty} dt \int_{\Omega}d\bm{r}\left|\left<\Psi_0\right|
\Psi(\bm{r},t)\left|\Phi\right>\right|^2.
\label{eq:dwell}
\end{equation}
The integral (\ref{eq:dwell})
converges since the probability of finding the scattered electron in $\Omega$
vanishes for time approaching $\pm \infty$.
After decomposing the wave packet on the basis made of $\left|\Psi_{\bm{k}}
\right>$'s and performing the time integration in (\ref{eq:dwell}), 
the delay time is obtained as a sum of stationary state
contributions \cite{Iannaccone95}, that are separately written as
\begin{eqnarray}
\tau_{\text{delay}}(\omega) &=&2\pi\hbar\,\delta\!(\hbar\omega-
\varepsilon_{\bm{k}})\times\nonumber\\
&&\int_{\Omega}\!\!d\bm{r}
\left[ \left|\left<\Psi_0\right|
\Psi(\bm{r})\left|\Psi_{\bm{k}}\right>\right|^2
- \left|\phi_{\bm{k}}\!(\bm{r})  \right|^2    \right].
\label{eq:delay2}
\end{eqnarray}
We extend the range of integration in (\ref{eq:delay2}) to the
whole space, since the contribution outside $\Omega$ is null, and
by combining (\ref{eq:delay}), (\ref{eq:delay2}), and the orbital
spectral representation of $\Psi(\bm{r})$, we obtain
the desired result
\begin{equation}
\frac{1}{\hbar}
\frac{d\, \delta_{\text{F}}(\omega)}{d \omega} = \pi\mathcal{N}(\omega),
\label{eq:super}
\end{equation}
where $\mathcal{N}(\omega)$ is the total spectral density
[cf.~(\ref{eq:N})]:
\begin{eqnarray}
\mathcal{N}(\omega) &=& -\frac{1}{\hbar\pi}\text{Im}\Big\{
\sum_{\alpha i}\mathcal{G}^{\alpha}_{ii}(\omega+\text{i}\eta) \nonumber\\
&+& \sum_{\bm k} \left[
\mathcal{G}_{\bm{k k}}(\omega+\text{i}\eta)
- G^{\,\text{free}}_{\bm{k k}}(\omega+\text{i}\eta)\right]\Big\}.
\label{eq:Ntot}
\end{eqnarray}

Equation (\ref{eq:super}) is the natural generalization
of (\ref{eq:Friedel}), namely a generalized FSR 
where the phase shift variation as a function of energy 
is proportional to the variation of
the total displaced {\em spectral} density $\mathcal{N}$. 
Again, all considerations of the previous 
example [Eq.~(\ref{eq:exemplum})] apply, 
i.e.~$\Delta \mathcal{N}$ does not need to be one between
two consecutive Coulomb blockade regions. The result (\ref{eq:super})
is generic and independent of the nature of the
coupling between dot and leads.

We are now able to focus on the experiment of Ref.~\onlinecite{Moty}.
The transmission phase shift of an electron tunneling into a QD 
is determined together with the integer charge filling in the dot, which
is in the Coulomb blockade regime. It turns out 
that the phase increment between specific neighboring
conductance valleys, as a function of the plunger voltage, 
is a fraction of $\pi$ ($\sim 0$ between
$N=3$ and $N=4$ and $\sim 3\pi/4$ between $N=6$ and $N=7$ blockaded regions).
Moreover, for these specific voltage ranges, the phase variation is 
continuous and smooth. Under such circumstances, $\delta_{\text{F}}$ and
the transmission phases are expected to coincide \cite{Lee99,Taniguchi99}.
According to previous theories based on the FSR \cite{Lee99,Taniguchi99}, 
one would predict a unit increment of $\pi$ in all cases, unless
some additional charge unpredictably accumulates outside the dot 
\cite{Yeyati95}. 
Our theory suggests an alternative natural explanation, 
i.e.~fractional (or even zero) increments of $\delta_{\text{F}}$ originate
from strong electron correlation (or spin blockade). This interpretation is
supported by the estimated low value of electron density. 
In fact, as the density diminishes, electrons in the dot are expected to  
crystallize \cite{Rontani06}, which affects $\delta_{\text{F}}$ in
a density-dependent manner. Extensive numerical simulations will be
reported elsewhere.

In conclusion, we derived an exact generalized Friedel sum rule for 
an interacting multi-level nano-object. The variation of the Friedel phase
through Coulomb blockade regions for values
which are fractions of $\pi$ is the fingerprint of electron correlation.

We thank E. Molinari, A. Ferretti, A. Calzolari, A. Bertoni, 
and G. Goldoni for stimulating 
discussions. This work is supported by MIUR-FIRB 
RBIN04EY74.

%

%\begin{thebibliography}{30}
%\bibliography{refbabbage}

\begin{thebibliography}{28}
\bibitem[{two()}]{twoleads}
\bibinfo{note}{The specialization of the formalism to two separate leads, as it
  is usually done in the QD literature, is straightforward but irrelevant in
  this context.}
\bibitem[{not()}]{notasimmetria}
\bibinfo{note}{Since most of the point-symmetry groups of realistic QDs are
  abelian, the $N_{\alpha}$ orbital levels belonging to the same irreducible
  representation $\alpha$ are generically non degenerate.}
\end{thebibliography}
%\end{thebibliography}

\end{document}